\documentclass[11pt,twoside]{article}
\usepackage[pdftex]{graphicx}
\usepackage{amsmath}
\usepackage{amssymb}
\usepackage{cite}

\newcommand{\bra}[1]{\left< #1 \right|}
\newcommand{\ket}[1]{\left| #1 \right>}

\newcommand{\eu}{\mathrm{e}}

\newcommand{\st}{^\star}

\newcommand{\half}{\frac{1}{2}}
\newcommand{\quat}{\frac{1}{4}}

\newcommand{\sqtwo}{{\sqrt{2}}}

\newcommand{\So}{\hat S}
\newcommand{\Xo}{{\hat X}}

\newcommand{\qo}{{\hat q}}
\newcommand{\po}{{\hat p}}
\newcommand{\co}{{\hat c}}
\newcommand{\cod}{{\hat c^\dagger}}
\newcommand{\Nor}{\mathcal{N}}
\newcommand{\Ord}{\mathcal{O}}
\newcommand{\pr}{^\prime}
\newcommand{\ka}{\kappa}
\newcommand{\coh}{\mathrm{ch}}
\newcommand{\tah}{\mathrm{th}}

\begin{document}
\newcommand{\pst}{\hspace*{1.5em}}

\newcommand{\rigmark}{\em Journal of Russian Laser Research}
\newcommand{\lemark}{\em Volume 30, Number 5, 2009}

\newcommand{\be}{\begin{equation}}
\newcommand{\ee}{\end{equation}}
\newcommand{\bm}{\boldmath}
\newcommand{\ds}{\displaystyle}
\newcommand{\bea}{\begin{eqnarray}}
\newcommand{\eea}{\end{eqnarray}}
\newcommand{\ba}{\begin{array}}
\newcommand{\ea}{\end{array}}

\thispagestyle{plain}

\label{sh}


\begin{center} {\Large \bf
\begin{tabular}{c}
NORMAL ORDERING THE SQUEEZE OPERATOR 
\\[-1mm]
BY GENERALIZED WICK THEOREM
\end{tabular}
 } \end{center}

\bigskip

\bigskip

\begin{center} {\bf
Lajos Di\'osi$^{*}$
}\end{center}

\medskip

\begin{center}
{\it
Wigner Research Centre for Physics\\
H-1525 Budapest 114, P.O.Box 49, Hungary
}
\smallskip

$^*$Corresponding author e-mail:~~~diosi.lajos@wigner.mta.hu\\
\end{center}

\begin{abstract}\noindent
In 1988, I translated a popular article \emph{Squeezed light} \cite{SluYur88} for
 the Hungarian mutation of Scientific American. 
How shall we say  \emph{squeezed} in our language from now on?
At the time, Janszky was already Hungary's top quantum optics theorist, 
he excelled in squeezed light research  as well \cite{Jan}, 
so I did not want to decide without asking him.
He voted for the word \emph{pr\'eselt} [prayshalt], and it worked  \cite{SluYur88Hun},
became standard in our talking and writing about squeezed light.
To his memory, let me dedicate an easy route (two pages) to normal 
ordering the squeeze operator, based on 
my recent extension of the Wick theorem for general
orderings of $q$ and $p$.   
\end{abstract}

\medskip


In quantum (as well as in classical) optics,  a single electromagnetic mode is equivalent with a harmonic oscillator whose dynamics is invariant for rotations 
of the phase plane spanned by coordinate  $\qo$ and momentum $\po$. 
The usual representation is the complex one, in terms of $\co=(\qo+i\po)/\sqrt{2}$
and its conjugate $\cod$, hence rotations simply become phase shifts of $\co$
and $\cod$. Squeezing $\qo$ into $\eu^{-r}\qo$ corresponds to the unitary
operator
\be\label{Scc}
\So=\exp\left(\frac{r}{2}(\co^2-\co^{\dagger 2})\right).
\ee
To confirm this, we can start from the elementary notion of squeezing: 
we have to rescale $\qo$ by a factor $1/\mu$ and  $\po$ by $\mu$. 
Re-write $\So$ in terms of $\qo,\po$:
\be\label{Sqp}
\So=\exp\left(\frac{i}{2}r \{\qo,\po\}\right)
\ee
yielding
\be
\So^\dagger\qo\So=\eu^{-r}\qo,~~~~\So^\dagger\po\So=\eu^{-r}\po
\ee
which is the desired squeezing with  $\mu=\eu^r$.

If we are interested in the coordinate representation
of $\So$, we have a definitive expression 
\cite{FanRua83}, as well a complementary one in momentum basis:  
\be\label{SFan}
\So=\frac{1}{\sqrt{\mu}}\int \ket{q/\mu}\bra{q}dq=\sqrt{\mu}\int\ket{\mu p}\bra{p}dp.
\ee
To see their equivalence with the standard forms (\ref{Scc},\ref{Sqp}),
let us derive each side. Both expressions of $\So$ satisfy the same simple 
differential equation: 
\be
\frac{d\So}{d\mu}=\left(-\frac{1}{2\mu}+\frac{i}{\mu}\po\qo\right)\So
                                  =\frac{i}{2\mu}\{\qo,\po\}\So.
\ee
With the initial condition $\So=\hat 1$ at $\mu=1$, we get the unique solution
coinciding with (\ref{Scc}) and (\ref{Sqp}).

For quantum optics, the normal ($\Nor$) ordered form of $\So$ is of
interest. It was derived e.g. in \cite{FanRua83,FanZaiKla87,Wun99,Fan03}. 
We shall nicely reduce the budget of calculations  if  we rely on the
generalized Wick theorem (GWT) \cite{Dio18} connecting two different orderings $\Ord,\Ord\pr$
of the exponential characteristic function $\exp(\Xo)$:
\be\label{GWT1}
\Ord\pr\eu^\Xo = \eu^C\Ord\eu^\Xo,
\ee
where $\Xo$ is any linear combination of $\qo,\po$ (or $\co,\cod$).
The pre-factor is a c-number because the exponent  is c-number:
\be\label{GWT2}
C=\half(\Ord\pr-\Ord)\Xo^2.
\ee
We call it the \emph{general contraction} between $\Ord\pr$ and $\Ord$.

To prepare the application of this theorem, consider the (first) expression of
$\So$ in (\ref{SFan}). For convenience, introduce  $\ka=1-1/\mu$ and insert 
the identity
\be
\ket{q/\mu}=\eu^{i\ka q\po}\ket{q}
\ee
into (\ref{SFan}). Observe that it leads to the PQ-ordered form:
\be\label{S1}
\So=\frac{1}{\sqrt{\mu}}\Ord_{PQ}\eu^{i(1-1/\mu)\po\qo},
\ee
where $\Ord_{PQ}$ puts the $\po$'s to the left of the $\qo$'s.
To apply GWT (\ref{GWT1}-\ref{GWT2}), we first unravel the 
bilinear form $\po\qo$ in the exponent. Let us insert the identity
\be
\eu^{i\ka\po\qo}=\int\eu^{iz\po\pm z\st\qo}\exp\left(-\frac{\vert z\vert^2}{\vert\ka\vert}\right)\frac{d^2z}{\pi\vert\ka\vert}
\ee
into (\ref{S1}), yielding
\be\label{S2}
{\hat S}=\int\Ord_{PQ}\eu^{iz\po\pm z\st\qo}\exp\left(-\frac{\vert z\vert^2}{\vert\ka\vert}\right)\frac{d^2z}{\pi\sqrt{\mu}\vert\ka\vert},
\ee
where the sign $\pm$ is the sign of $\ka$. Now we can apply our GWT (\ref{GWT1}-\ref{GWT2}):
\bea\label{S3}
\Ord_{PQ}\eu^{iz\po\pm z\st\qo}&=&\eu^C \Nor\eu^{iz\po\pm z\st\qo},\\
C&=&\half(\Ord_{PQ}-\Nor)(iz\po\pm z\st\qo)^2.
\eea
Using the relations 
\begin{eqnarray}
\Nor\qo^2&\equiv&\Nor\frac{(\co+\cod)^2}{2}=\qo^2-\half\nonumber\\
\Nor\po^2&\equiv&\Nor\frac{(\co-\cod)^2}{2i}=\po^2-\half\\
\Nor\qo\po&=&\half\{\qo,\po\}\nonumber
\end{eqnarray}
together with $\Ord_{PQ}\qo\po=\Ord_{PQ}\po\qo=\po\qo$,
we get the c-number contraction:
\be\label{Czz}
C=\quat(z^{\star 2}-z^2)\pm\half\vert z\vert^2.
\ee
Now (\ref{S2}) takes this form:
\be\label{}
\So\!=\!\Nor\!\!\int\!\!\exp
\!\left(\!
iz\po\pm z\st\qo+\frac{z^{\star 2}-z^2}{4}-\left\vert\frac{1}{\ka}-\half\right\vert \vert z\vert^2
\!\right)\!
\frac{d^2z}{\pi\sqrt{\mu}\vert\ka\vert}.
\ee
Thanks to GWT, we only needed elementary calculational steps so far, and 
the evaluation of the  Gaussian integral yields the desired normal ordered squeeze
operator:
\be\label{SqpN}
\So\!=\!\sqrt{\!\frac{2\mu}{1\!+\!\mu^2}}\Nor\!\exp\!\left\{\frac{i(\mu^2\!-\!1)\po\qo\!-\!\half(\mu\!-\!1)^2(\po^2\!+\!\qo^2)}{1+\mu^2}\right\},
\ee
where we expressed $\ka$ as $1-1/\mu$.

It is worthwhile to notice that, apart from the powerful GWT \cite{Dio18}, which is a 
consequence of the Baker--Campbell--Haussdorff theorem \cite{Cam97,Bak02,Hau06} 
so common in
quantum optics, our method is akin to Fan's ``integration within ordered products''
\cite{FanRua83,FanZaiKla87,Wun99,Fan03}. Also the technical formula 
\be
\int\exp(\zeta\vert z\vert^2+\xi z +\eta z\st +f z^2 + g z^{\star 2})\frac{d^2z}{\pi}
=\frac{1}{\sqrt{\zeta^2-4fg}}\frac{-\zeta\xi\eta+\xi^2g+\eta^2f}{\zeta^2-4fg}
\ee 
 to evaluate our (\ref{S2}) can be borrowed from \cite{Fan03}. 

The natural representation of normal ordered expressions is in terms
$\co$ and $\cod$, of course, rather than in $\qo$ and $\po$. Let us rewrite (\ref{SqpN})
accordingly:
\be
\So\!=\!\sqrt{\!\frac{2\mu}{1\!+\!\mu^2}}\Nor\!\exp\left\{\frac{\half(\mu^2\!-\!1)(\co^2\!-\!\co^{\dagger2})\!+\!(\mu\!-\!1)^2\cod\co}{\mu^2+1}\right\}
\ee
which, using the standard squeezing parameter $r$, takes this form:
\be
\So\!=\!\frac{1}{\sqrt{\coh r}}\Nor\!\exp\left\{\half\tah r(\co^2\!-\!\co^{\dagger2})\!+\!(\frac{1}{\coh r} -1)\cod\co\right\}
\!=\!\frac{1}{\sqrt{\coh r}}\exp(-\half\tah r \co^{\dagger2})
\left(\frac{1}{\coh r}\right)^{\!\!\cod\co}
\!\!\exp(\half\tah r \co^2)
\ee
coinciding with (8.13) of \cite{Wun99} --- as it should.

This short derivation, an expansion of the example already outlined in \cite{Dio18},
gives us the opportunity to visualize how GWT reduces the calculational budget
in general. We start from the PQ-ordered l.h.s. of (\ref{S3}) (with notations $a=z\st,b=iz$) 
and normal order it without using the GWT:
\begin{eqnarray}
\Ord_{PQ}\eu^{a\qo+b\po}&=&\eu^{(b/i\sqtwo)(\co-\cod)}\eu^{(a/\sqtwo)(\co+\cod)}
=\eu^{(a^2+b^2)/4}\eu^{-(b/i\sqtwo)\cod}\eu^{(b/i\sqtwo)\co}
        \eu^{a\sqtwo)\cod}\eu^{(a\sqtwo)\co}=\nonumber\\
&=&\eu^{(a^2+b^2)/4-(i/2)ab}\eu^{-(b/i\sqtwo)\cod}\eu^{a\sqtwo)\cod}\eu^{(b/i\sqtwo)\co}\eu^{(a\sqtwo)\co}
=\eu^{(a^2+b^2)/4-(i/2)ab}\Nor\eu^{a\qo+b\po}.   
\end{eqnarray}
Here the Baker--Campbell--Haussdorff theorem applies three times step-by-step. 
This can be spared by applying the GWT in a single step, as we did before.

\section*{Acknowledgments}
\pst
The author thanks the National Research Development and
Innovation Office of Hungary projects 2017-1.2.1-NKP-2017-00001 and K12435,
and the EU COST Action CA15220 for support.

\end{document}